\documentclass[preprint,amsmath,amssymb,prl]{revtex4}

\usepackage{mathrsfs}
\usepackage{graphicx,color}
\usepackage{dcolumn}
\usepackage{bm}
\usepackage{CJK}
\usepackage{epstopdf}
\usepackage[colorlinks=true,urlcolor=blue,linkcolor=blue,citecolor=blue]{hyperref}

\begin{document}
\begin{CJK*}{UTF8}{gbsn}
\title{Nuclear mass predictions with machine learning reaching the accuracy required by $r$-process studies}

\author{Z. M. Niu$^{1}$}\email{zmniu@ahu.edu.cn}
\author{H. Z. Liang$^{2,3}$}\email{haozhao.liang@phys.s.u-tokyo.ac.jp}

\affiliation{$^1$School of Physics and Optoelectronic Engineering, Anhui University,
             Hefei 230601, China}
\affiliation{$^2$Department of Physics, Graduate School of Science, The University of Tokyo,
             Tokyo 113-0033, Japan}
\affiliation{$^3$RIKEN Nishina Center, Wako 351-0198, Japan}

\date{\today}

\begin{abstract}
Nuclear masses are predicted with the Bayesian neural networks by learning the mass surface of even-even nuclei and the correlation energies to their neighbouring nuclei. By keeping the known physics in various sophisticated mass models and performing the delicate design of neural networks, the proposed Bayesian machine learning (BML) mass model achieves an accuracy of $84$~keV, which crosses the accuracy threshold of the $100$~keV in the experimentally known region. It is also demonstrated the corresponding uncertainties of mass predictions are properly evaluated, while the uncertainties increase by about $50$~keV each step along the isotopic chains towards the unknown region. The shell structures in the known region are well described and several important features in the unknown region are predicted, such as the new magic numbers around $N = 40$, the robustness of $N = 82$ shell, the quenching of $N = 126$ shell, and the smooth separation energies around $N = 104$.
\end{abstract}

\maketitle

\textit{Introduction}---The origin of heavy elements in the Universe is an important but unanswered fundamental question of science~\cite{Haseltine2002Discover}. The rapid neutron-capture process ($r$-process) is responsible for producing about half of the elements heavier than iron~\cite{Burbidge1957RMP}. During the past decades, the $r$-process studies have made substantial progress from both nuclear physics and astrophysics sides~\cite{Arnould2007PRp, Cowan2021RMP}. However, the $r$-process astrophysical sources and their specific conditions remain mysteries, and the identification of the most important $r$-process site also remains a hot topic~\cite{Smartt2017Nature, Watson2019Nature, Siegel2019Nature}.

The $r$-process studies necessitate the joint efforts of nuclear physicists and astrophysicists~\cite{Kajino2019PPNP}. From the nuclear side, nuclear mass is a crucial input~\cite{Martin2016PRL}, which determines the $r$-process path, and hence relates the main $r$-process abundance peaks at $A=130$ and $195$ to the nuclear shell closures at $N=82$ and $126$, respectively. Nuclear mass also determines the reaction energies of $\beta$ decay and neutron capture in the $r$ process, so it is one important source of theoretical uncertainties of $\beta$-decay half-lives and neutron-capture rates~\cite{Li2019SCPMA, Ma2019PRC}. Although the measurements of nuclear mass have been made great progress in recent years, especially for the nuclei on the $r$-process path around $N=82$~\cite{Wang2021CPC}, the $r$-process path near $N=126$ or above is still unreachable for the present, or even the next-generation, radioactive ion beam facilities. Therefore, accurate nuclear mass predictions are essential to understand the mysteries in the $r$ process.

Due to the difficulties in the quantum many-body problem and the complexity of nuclear force, accurate nuclear mass prediction is a very challenging theoretical task. Even in the experimentally known region, the accuracies of nuclear mass predictions are generally around $500$ keV~\cite{Lunney2003RMP}, which is much poorer than the accuracy of $100$ keV required by the $r$-process studies~\cite{Mumpower2016PPNP}. The greater difficulty lies in the extrapolation. It is found that the deviations of different mass models can even reach tens of MeV when they are extrapolated to the unknown neutron-drip line. Therefore, the accurate nuclear mass prediction has become one of the bottlenecks in the $r$-process studies.

In particular, one of the hot topics in the $r$-process studies during past decades is the origin of the rare-earth peak, which has been claimed to be associated with the $N\approx 104$ kink in the separation energies~\cite{Surman1997PRL} or the doubly asymmetric fission fragment distributions in the $A\approx 278$ region~\cite{Goriely2013PRL}. If one can construct accurate enough mass predictions for the $r$-path nuclei leading to the rare-earth peak, one can confirm whether there is a kink in the separation energies near $N=104$, which will become an essential step for understanding the origin of the rare-earth peak.

For the above key open questions, we recall that the famous Bethe-Weizs\"{a}cker (BW) formula is the first nuclear mass model, in which the nucleus is assumed as a charged liquid drop~\cite{Weizsacker1935ZP, Bethe1936RMP}. It achieves an accuracy of about $3$~MeV, while large deviations from the experimental data are found in the nuclei near the magic numbers. These large deviations can be reduced by including the microscopic correction energies, and the nuclear mass predictions with the accuracy of about $300$--$500$~keV can be obtained. This kind of mass model is usually named as ``macroscopic-microscopic'' model, such as finite-range droplet model (FRDM)~\cite{Moller2012PRL} and Weizs\"{a}cker-Skyrme (WS) model~\cite{Wang2014PLB}. However, the microscopic correction energies are generally extracted from the single-particle levels of phenomenological mean fields, which are generally independent of the macroscopic part. Such an inconsistency between the macroscopic and microscopic parts would affect the model reliability. The microscopic mass models based on the nuclear density functional theory are usually believed to have better extrapolation abilities, e.g., the relativistic mean-field model~\cite{Geng2005PTP, Xia2018ADNDT} and the nonrelativistic Hartree-Fock-Bogliubov (HFB) model with Skyrme~\cite{Goriely2009PRLa} or Gogny~\cite{Goriely2009PRLb} force. Their present accuracies are, however, generally lower than the macroscopic-microscopic models.

To further improve the accuracy of nuclear models, the machine learning techniques have attracted much attention during the past years. In particular, the Bayesian version of machine learning is expected to be able to provide the corresponding theoretical uncertainties~\cite{Utama2016PRC}. For the nuclear mass predictions with Bayesian neuron network (BNN), we pointed out that the performance of BNN can be improved by enriching the network inputs with information of physics~\cite{Niu2018PLB}, such as the pairing and shell effects. Neufcourt \emph{et al}.~\cite{Neufcourt2018PRC} agreed with this idea in their study of two-neutron separation energies. Since then, nuclear structure with machine learning techniques has become a hot frontier, for example, in the studies of neutron-drip line in the Ca region~\cite{Neufcourt2019PRL}, the incomplete fission yields~\cite{Wang2019PRL}, and the low-lying excitation spectra~\cite{Lasseri2020PRL}. From the above studies, see also a recent review~\cite{Bedaque2021EPJA} and the references therein, one can conclude that the accuracy and the capability of extrapolation of study with machine learning techniques crucially depend on the delicate designs of neuron network, by taking into account as much physics as possible.

In this Letter, we propose a nuclear mass model with Bayesian machine learning and pay special attention on the designs of the structure, outputs, and inputs of the neuron networks. We will first demonstrate the accuracy of mass prediction as well as the capability of extrapolation of the present model with a theory-to-theory validation. We will then show the present BML mass model achieves an accuracy of $84$~keV with respect to the experimental data in AME2016~\cite{Wang2017CPC} and also discuss the shell structures in the experimentally known and unknown regions, which are crucial for the $r$-process studies.

\textit{Designs of BNN}---In the present study, we adopt the general scheme of BNN~\cite{Neal1996Book}. BNN can avoid the over-fitting problem automatically by using the hyper priors. It can also quantify the uncertainties in predictions, since all model parameters are described with probability distributions.

For the present designs of the network structure, we keep in mind that the physics (e.g., the ground-state spin and parity) of odd-$A$ and odd-odd nuclei are much more sophisticated than that of even-even nuclei. Thus, the predictive power, especially the extrapolation capability, will be substantially affected if we directly train the neural network with the whole nuclear mass surface. A much more effective strategy is the training of neural network with the smoother mass surface of even-even nuclei, together with the trainings with the separation energies related to their neighbouring odd-$A$ and odd-odd nuclei. As a result, there are in total $9$ different BNNs to cover the mass predictions for the whole nuclear chart. See Fig.~\textcolor[rgb]{0.00,0.00,1.00}{1}~in Supplemental Materials~\cite{SuppMat} and the corresponding descriptions.

For the designs of the network outputs, in our previous study~\cite{Niu2018SciB}, we showed quantitatively that the performance of machine learning is very limited if crucial information of physics is missing. The discrepancy between the experimental data and the predictions of a given model $\delta M = M^{\rm exp} - M^{\rm model} $ is usually taken as the output, i.e., the learning target \cite{Utama2016PRC, Niu2018PLB}, which can effectively inlcude the known physics in the given model. To make the best use of the established nuclear mass models, we employ the macroscopic model BW2~\cite{Kirson2008NPA}, the macroscopic-microscopic models KTUY~\cite{Koura2005PTP}, FRDM12~\cite{Moller2012PRL}, and WS4~\cite{Wang2014PLB}, the microscopic models RMF~\cite{Geng2005PTP} and HFB-31~\cite{Goriely2016PRC}, and other high-precision global mass models Bhagwat~\cite{Bhagwat2014PRC} and DZ28~\cite{Duflo1995PRC}. These mass models have taken into account the physics important to the description of nuclear mass from different aspects.

For the designs of the network inputs, in addition to $Z$ and $N$, we further introduce $E^{\rm model}_{\rm mic} \equiv M^{\rm model} - E_{\rm mac}$ or the counterparts of the separation energies as an input. This quantity is completely missing in the macroscopic mass models, while it is related to the effective mass of nucleon in the microscopic mass models. It can be seen that the prefect mass model that reproduces all the experimental data holds a prefect correlation between the input and output, $E^{\rm model}_{\rm mic} = E_{\rm mic}$, independent of $Z$ and $N$. In such a way, the systematic overestimation or underestimation on $E^{\rm model}_{\rm mic}$ of a given model can be corrected by BNN in an efficient way. In principle, $E_{\rm mac}$ can be taken as any smooth function of $Z$ and $N$ on the nuclear mass surface. Here, it is taken from the macroscopic part of FRDM12.

Based on each mass model $i$, we can get its corresponding BNN mass prediction $M_i$ with the error $\sigma_i$. To describe the systematic error of mass prediction, the weighted mean $M$ and standard deviation $\sigma_{M}$ of $M_i$ are taken as the final mass prediction, which are
\begin{eqnarray}\label{Eq:Maver}
  M = \frac{\sum_{i=1}^m \omega_i M_i}{\sum_{i=1}^m \omega_i},
  \quad
  \sigma_{M} = \frac{1}{\sqrt{\sum_{i=1}^m \omega_i}}
\end{eqnarray}
where $\omega_i=1/\sigma_i^2$ and $m$ is the number of mass models. Since some sources of error may not be taken into account, the error $\sigma_{M}$ is further corrected with a factor $\chi_\nu$, which considers the deviations between mass predictions $M$ and experimental data
\begin{eqnarray}
  \chi_\nu^2 = \frac1n \sum_{Z,N\geqslant 8} \frac{1}{\sigma_{M(Z,N)}^2}
              \left[M(Z,N) - M_{\rm exp}(Z,N)\right]^2,
\end{eqnarray}
where $n$ is the number of nuclei in the learning set. Here $\chi_\nu=2.1$ for the experimental data in AME2016. For simplicity, this Bayesian machine learning model described above will be denoted by BML hereafter.

Before ending this part, we perform a theory-to-theory validation with the above designs of BNN, to demonstrate the accuracy of prediction and the capability of extrapolation. In such a benchmark calculation, the nuclear masses of FRDM12 are used as the pseudo experimental data (i.e., the target values). Meanwhile, the other 7 mass models---BW2, KTUY, WS4, RMF, HFB-31, Bhagwat, and DZ28---are regarded as our present knowledge and used as the inputs of BNN. To simulate the present experimentally known region, the learning set is limited to those nuclei listed in AME2016, and all nuclei outside AME2016 will be used to testify the extrapolation capability. As a result, the mass prediction accuracy in the learning region reaches $93$~keV. It is also found that the mass prediction uncertainties increase by about $50$~keV each step along the isotopic chains towards the unknown region, which agrees with the standard deviations between the mass predictions and the corresponding FRDM12 values. For details, see Fig.~\textcolor[rgb]{0.00,0.00,1.00}{2}~in Supplemental Materials~\cite{SuppMat} and the corresponding discussion.

\begin{figure}
\includegraphics[width=8cm]{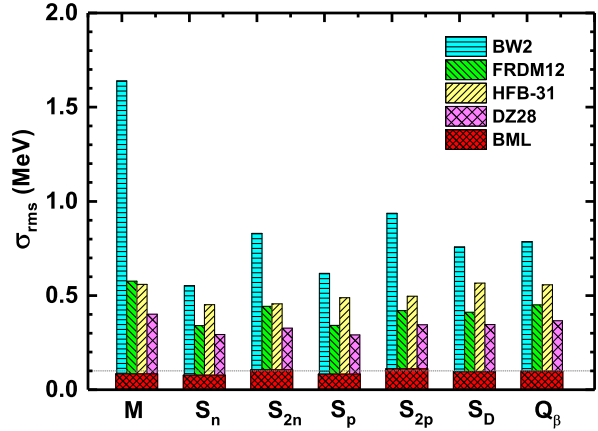}
\caption{(Color online) The rms deviations of $M$, $S_n$, $S_{2n}$, $S_p$, $S_{2p}$, $S_D$, and $Q_{\beta}$ with respect to the experimental data for the learning set of the BML mass model.
The corresponding rms deviations given by the BW2, FRDM12, HFB-31, and DZ28 mass models are shown for comparison.}
\label{Fig:sigrms}
\end{figure}

\textit{Results and Discussion}---Using the high-precision experimental data in AME2016 as the learning set, we construct the mass predictions of the present BML model. The root-mean-square (rms) deviations of $M$ and various separation or decay energies with respect to the experimental data for the learning set are given in Fig.~\ref{Fig:sigrms}. For comparison, the corresponding rms deviations given by some other mass models are also given. It is clear that the BML model achieves a very high accuracy of mass prediction, which is of the best accuracy for global mass predictions as we have known and for the first time crosses the accuracy threshold of $100$~keV in the known region. Furthermore, the BML model also achieves high accuracies for various separation or decay energies, which are at least about $3$ times higher than other shown mass models. Even comparing with the previous machine learning model WS4+BNN-I4~\cite{Niu2018PLB}, whose corresponding rms values are $184$, $208$, $216$, $213$, $227$, and $255$~keV for mass, $S_n$, $S_p$, $S_{2n}$, $S_{2p}$, and $Q_{\beta}$, respectively, the present BML model achieves much smaller rms values, i.e., $84$, $78$, $83$, $105$, $111$, and $99$~keV, respectively. This indicates the BML model describes excellently not only the mass surface globally but also its local details, including its derivatives in different directions on the nuclear chart.

\begin{figure}
\includegraphics[width=8cm]{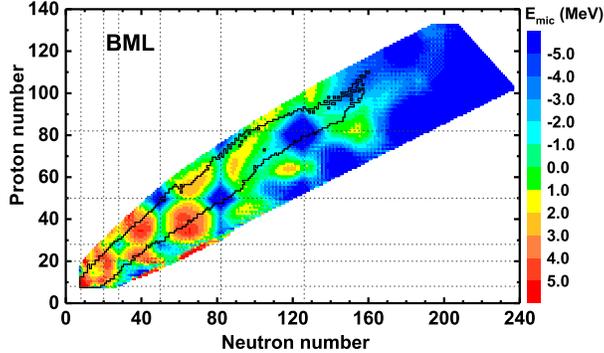}
\caption{(Color online) Microscopic correction energies $E_{\rm mic}$ of BML.
The contours show the boundary of nuclei with known masses in AME2016 and the dotted lines denote the traditional magic numbers.}
\label{Fig:Emic}
\end{figure}

The microscopic correction energies $E_{\rm mic}$ can reveal the shell effects in nuclear properties. Therefore, we show $E_{\rm mic}^{\rm BML} = E^{\rm BML} - E_{\rm mac}$ of BML in Fig.~\ref{Fig:Emic}. It is clear that the shell structures in the known region are well reproduced. Being extrapolated to the unknown region, even to the drip lines, there are several remarkable structure features, which are hardly achieved by other learning approaches, such as the radial basis function approach~\cite{Niu2013PRCb, Niu2016PRC, Niu2019PRCb}. Apart from the traditional magic numbers, the new magic numbers around $N=40$ in the light nuclei region and those around $Z=120$ in the superheavy nuclei region are also predicted by BML.

It is well known that $N=82$ and $N=126$ shells are crucial for the $r$-process properties, e.g., they are responsible for the main peaks of solar $r$-abundance at $A=130$ and $A=195$, respectively. From Fig.~\ref{Fig:Emic}, it is found that the $N=82$ shell remains robust even going to the neutron-drip line, which has been approved by recent experimental studies~\cite{Watanabe2013PRL}. However, we predict that the $N=126$ shell will first quench and then enhance as approaching the proton magic number $Z=50$ when going to the neutron-drip line. The $N=126$ shell also quenches when going to the proton-drip line, even just away from the known region.

\begin{figure}
\includegraphics[width=8cm]{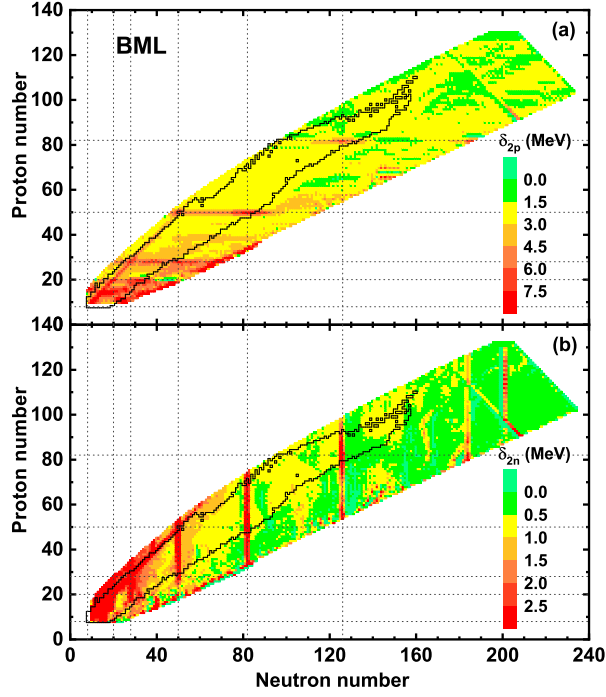}
\caption{(Color online) Same as Fig.~\ref{Fig:Emic} but for (a) the two-proton gaps $\delta_{2p}$ and (b) the two-neutron gaps $\delta_{2n}$.}
\label{Fig:d2nd2p}
\end{figure}

The two-proton (neutron) gaps $\delta_{2p}$ ($\delta_{2n}$) are also important signatures of nuclear magic numbers, which take local maxima at proton (neutron) magic numbers. From Fig.~\ref{Fig:d2nd2p}, which shows $\delta_{2p}$ and $\delta_{2n}$ of BML, the traditional magic numbers are well exhibited. The BML model predicts a neutron magic number at $N=184$, although its $\delta_{2n}$ is not as strong as those of traditional magic numbers. It should be pointed out that the larger $\delta_{2n}$ at $N\thickapprox 200$ is not necessarily a signature of magic number, which mainly originates from the lack of mass predictions for nuclei with $N>200$ in the KTUY model.

\begin{figure*}
\includegraphics[width=14cm]{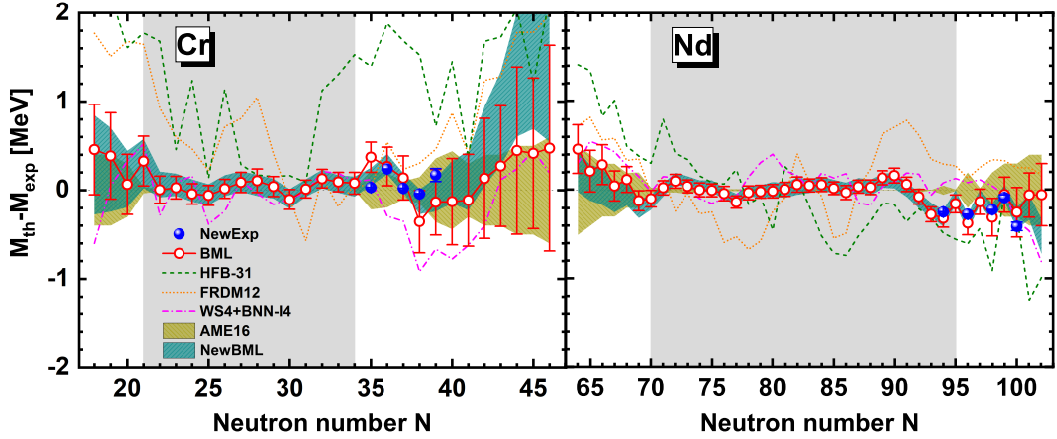}
\caption{(Color online) Mass differences between $M_{\rm th}$ from the BML, NewBML, HFB-31, FRDM12, and WS4+BNN-I4 mass models and $M_{\rm exp}$ in AME2016 for the Cr and Nd isotopes. The shaded (white) regions indicate the BML learning (extrapolation) areas, where the accuracy of $M_{\rm exp}$ in AME2016 is higher (worse) than $100$~keV. The new experimental data~\cite{Mougeot2018PRL, Orford2018PRL} after AME2016 are shown with blue solid symbols.}
\label{Fig:MdCrNd}
\end{figure*}

To show the details of the present BML mass model and illustrate explicitly its extrapolation capability, the mass differences between $M_{\rm th}$ from various mass models and $M_{\rm exp}$ in AME2016 are shown in Fig.~\ref{Fig:MdCrNd}, by taking the Cr and Nd isotopes as examples. In particular, in these two isotopic chains, there are several experimental data on both neutron-rich and proton-rich sides with the accuracy worse than $100$~keV, shown as the white regions in Fig.~\ref{Fig:MdCrNd}. Therefore, we did not include those data in the learning set.

It is clear that in the BML learning areas, the shaded regions in Fig.~\ref{Fig:MdCrNd}, the BML mass predictions are in an excellent agreement with the experimental data with an accuracy around $100$~keV, apart from the region around $^{154}$Nd. It is also seen that in the extrapolation areas, both neutron-rich and proton-rich sides, the BML mass predictions agree with the experimental data within the experimental and theoretical uncertainties. Remarkably, we still hold such a nice agreement, when we extrapolate the mass predictions from $^{58}$Cr to $^{70}$Cr with $12$ neutrons more. In both learning and extrapolation areas, the performance of BML is much better than those of HFB-31 and FRDM12, even better than the previous machine learning results of WS4+BNN-I4.

In the regions of Cr and Nd, there are a number of new experimental data~\cite{Mougeot2018PRL, Orford2018PRL} after AME2016, which are shown with blue solid symbols in Fig.~\ref{Fig:MdCrNd}. The comparison between the new data and the BML mass predictions again show excellent agreements not only on the values of the mass but also on the systematics of the mass surface. For example, the BML mass prediction on $^{154}$Nd is consistent with the new data, instead of that in AME2016.

As a step further, to show the influence of new experimental data, the new BML mass predictions are made by including these new data after AME2016~\cite{Mougeot2018PRL, Orford2018PRL, Vilen2018PRL, Vilen2019PRC, Vilen2020PRC, Canete2020PRC, Welker2017PRL, Manea2020PRL, Leistenschneider2018PRL, Reiter2020PRC, Ito2018PRL, Michimasa2018PRL, Izzo2018PRC, Ong2018PRC, Valverde2018PRL, Puentes2020PRC, Althubiti2017PRC, Ascher2019PRC, Nesterenko2017JPG, Brodeur2017PRC, Reiter2018PRC, Babcock2018PRC, Hartley2018PRL, Zhang2018PRC, Xu2019PRCR, Xu2019PRC, Andres2020EPJA, Mougeot2020PRC} into the learning set, which are denoted by NewBML for simplicity. The corresponding results are also shown in Fig.~\ref{Fig:MdCrNd} with dark-green shaded bands. It is found that, if the new data are included in the learning set, the theoretical uncertainties near the new data reduce to about half of the original values.

\begin{figure}
\includegraphics[width=8cm]{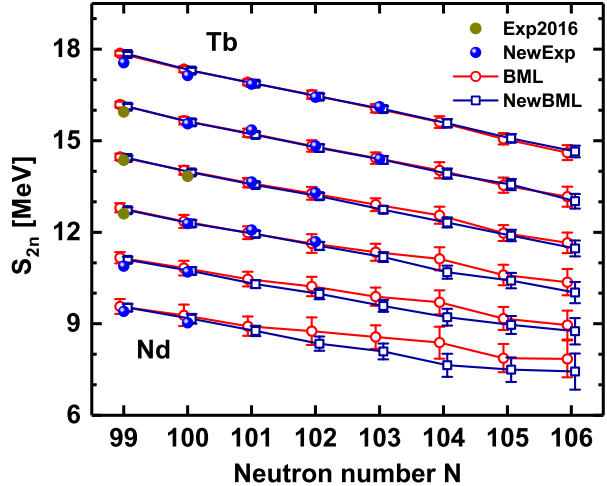}
\caption{(Color online) Two-neutron separation energies $S_{2n}$ of $Z=60$--$65$ isotopes. The $S_{2n}$ calculated with the experimental data in AME2016 are shown with yellow circles, while the new $S_{2n}$ calculated by including the new experimental data~\cite{Mougeot2018PRL, Orford2018PRL, Vilen2018PRL, Vilen2019PRC, Vilen2020PRC, Canete2020PRC, Welker2017PRL, Manea2020PRL, Leistenschneider2018PRL, Reiter2020PRC, Ito2018PRL, Michimasa2018PRL, Izzo2018PRC, Ong2018PRC, Valverde2018PRL, Puentes2020PRC, Althubiti2017PRC, Ascher2019PRC, Nesterenko2017JPG, Brodeur2017PRC, Reiter2018PRC, Babcock2018PRC, Hartley2018PRL, Zhang2018PRC, Xu2019PRCR, Xu2019PRC, Andres2020EPJA, Mougeot2020PRC} are shown with blue circles. The $S_{2n}$ predictions of BML and NewBML are shown with open circles and open squares, respectively. For displaying the data clearly, all $S_{2n}$ of $Z=61$--$65$ isotopes are increased by $(Z-60)$ MeV.}
\label{Fig:S2nInfNewExp}
\end{figure}

For the important issue related to the origin of the rare-earth peak and the possible kinks in the separation energies near $N = 104$, we show in Fig.~\ref{Fig:S2nInfNewExp} the two-neutron separation energies $S_{2n}$ for the $Z = 60$--$65$ isotopes. While the BML mass predictions agree well with both AME2016 and new data in this region, the new data can further substantially reduce the theoretical uncertainties of $S_{2n}$ for the neighboring nuclei. As a result, the $S_{2n}$ predictions around $N=104$ by NewBML tend to be smooth, rather than with kinks. In other words, it is more likely that the origin of the rare-earth peak is due to the doubly asymmetric fission fragment distributions in the $A\approx 278$ region~\cite{Goriely2013PRL}. More experimental data in the coming years will further testify this conclusion. By taking the new nuclei with $Z, N \geqslant 8$ first appearing in latest database AME2020~\cite{Wang2021CPC} as the testing set, the rms deviation of BML model with respect to those new data with the experimental uncertainties smaller than $100$~keV is $170$~keV. This indicates a good accuracy is also achieved by the BML model for these new data, which are not in the training. In contrast, the corresponding rms deviations are $245$, $691$, and $718$~keV for WS4+BNN-I4~\cite{Niu2018PLB}, FRDM2012~\cite{Moller2012PRL}, and HFB-31~\cite{Goriely2016PRC} models, respectively.

Finally, all experimental masses with $Z, N \geqslant 8$ and uncertainties smaller than $100$ keV in the latest database AME2020~\cite{Wang2021CPC} are employed to train the BML model, the resulting mass predictions are given in the Supplemental Materials~\cite{SuppMat}.

\textit{Summary}---High-precision mass predictions are made with the Bayesian neural networks by learning the mass surface of even-even nuclei and the correlation energies to their neighbouring nuclei. The known physics in various mass models are kept to achieve good predictive capability. With this strategy, the proposed BML mass model describes well not only the mass surface globally but also its local details including its derivatives in different directions on the nuclear chart. As a result, BML achieves high accuracy for both nuclear masses and various separation or decay energies. The accuracy of BML mass predictions reaches $84$~keV, which has crossed the accuracy threshold of $100$~keV in the known region. The uncertainties of BML mass predictions are also reasonably evaluated, which increase about $50$~keV as going forward one step along the isotopic chain from the known region, and the new experimental data after AME2016 can be precisely predicted by BML within the experimental and theoretical uncertainties.

While the shell structures in the known region are well described, we also predict several important features in the unknown region, such as the new magic numbers around $N = 40$, the robustness of $N = 82$ shell, the quenching of $N = 126$ shell, and the smooth separation energies around $N = 104$, which are all crucial for the quantitative $r$-process calculations.

With the present designs of the BML mass model, the future experimental data of nuclear mass as well as the future advanced nuclear mass models can be taken into account by the same strategy. The nuclear mass predictions towards the unknown region can be carried out and improved systematically and continuously.

\section*{Acknowledgements}
We are grateful to Professor Shan-Gui Zhou and Professor Yi-Fei Niu for the fruitful discussions. This work was partly supported by the National Natural Science Foundation of China under Grant No.~11875070 and No.~11935001, the Anhui project (Z010118169), the JSPS Grant-in-Aid for Early-Career Scientists under Grant No.~18K13549, the JSPS Grant-in-Aid for Scientific Research (S) under Grant No.~20H05648, the RIKEN iTHEMS program, and the RIKEN Pioneering Project: Evolution of Matter in the Universe. The authors acknowledge the High-performance Computing Platform of Anhui University for providing computing resources.



\end{CJK*}
\end{document}